\documentclass[10pt,draft]{article}
\usepackage{amsfonts,amsmath,amssymb}
\newcommand{\Real}{\mathop{\textrm{Re}}}
\begin{document}
\title{Comment on ``Four-component relativistic theory for NMR
parameters: Unified formulation and numerical assessment of different
approaches'' [J.\ Chem.\ Phys.\ 130, 144102 (2009)]}
\author{Rados{\l}aw Szmytkowski\footnote{Corresponding author. 
Email: radek@mif.pg.gda.pl}\hspace*{1ex}
and
Patrycja Stefa{\'n}ska \\*[3ex]
Atomic Physics Division,
Department of Atomic Physics and Luminescence, \\
Faculty of Applied Physics and Mathematics,
Gda{\'n}sk University of Technology, \\
Narutowicza 11/12, 80--233 Gda{\'n}sk, Poland}
\date{\today}
\maketitle
\begin{abstract}
In the paper commented on [J.\ Chem.\ Phys.\ 130 (2009) 144102],
Cheng \emph{et al.\/} derived a formula for the magnetic dipole
shielding constant $\sigma$ for the Dirac one-electron atom in its
ground state. That formula involves an infinite series of ratios of
the Euler's gamma functions. We show that with some algebra the
series may be expressed in terms of elementary functions. This leads
to a simple closed-form expression for the shielding constant.
\end{abstract}
\section*{}
In a recent paper \cite{Chen09}, Cheng \emph{et al.\/} have discussed
a four-component relativistic theory of NMR parameters. In
particular, they have applied an analytical calculation technique,
based on the Sturmian expansion of the first-order Dirac--Coulomb
Green function, found some years ago by one of us in Ref.\
\cite{Szmy97}, to derive a formula for the magnetic dipole shielding
constant $\sigma$ for the Dirac one-electron atom in its ground
state. They have shown that $\sigma$ may be written in the form
\begin{equation}
\sigma=\sigma_{-1}+\sigma_{+2},
\label{1}
\end{equation}
where
\begin{equation}
\sigma_{-1}=-\frac{2Z\alpha^{2}}{9}
\frac{2\gamma_{1}^{2}+\gamma_{1}-4}{\gamma_{1}(2\gamma_{1}-1)}
\label{2}
\end{equation}
and
\begin{equation}
\sigma_{+2}=\frac{2Z\alpha^{2}}{9}
\frac{\Gamma(\gamma_{2}+\gamma_{1}-1)\Gamma(\gamma_{2}+\gamma_{1}+2)}
{\Gamma(\gamma_{2}-\gamma_{1}-1)\Gamma(\gamma_{2}-\gamma_{1}+2)
\Gamma(2\gamma_{1}+1)}
\sum_{n=0}^{\infty}\frac{\Gamma(n+\gamma_{2}-\gamma_{1}-1)
\Gamma(n+\gamma_{2}-\gamma_{1}+2)}{n!\Gamma(n+2\gamma_{2}+1)
(n+\gamma_{2}-\gamma_{1})},
\label{3}
\end{equation}
with
\begin{equation}
\gamma_{\kappa}=\sqrt{\kappa^{2}-(\alpha Z)^{2}}.
\label{4}
\end{equation}
Here $Z$ is the nuclear charge, $\alpha$ is the Sommerfeld's
fine-structure constant, while $\Gamma(\zeta)$ is the Euler's gamma
function.

The representation of $\sigma_{+2}$ displayed in Eq.\ (\ref{3}) looks
formidable. It is the purpose of this comment to prove that the
series in Eq.\ (\ref{3}) may be summed to a closed form, leading to
an extremely simple expression for $\sigma_{+2}$.

To begin, we observe that with the aid of the well-known property
$\zeta\Gamma(\zeta)=\Gamma(\zeta+1)$, we may write
\begin{equation}
\Gamma(n+\gamma_{2}-\gamma_{1}+2)
=(n+\gamma_{2}-\gamma_{1})[\Gamma(n+\gamma_{2}-\gamma_{1})
+\Gamma(n+\gamma_{2}-\gamma_{1}+1)].
\label{5}
\end{equation}
Plugging Eq.\ (\ref{5}) into Eq.\ (\ref{3}) yields
\begin{eqnarray}
\sigma_{+2} &=& \frac{2Z\alpha^{2}}{9}
\frac{\Gamma(\gamma_{2}+\gamma_{1}-1)\Gamma(\gamma_{2}+\gamma_{1}+2)}
{\Gamma(\gamma_{2}-\gamma_{1}-1)\Gamma(\gamma_{2}-\gamma_{1}+2)
\Gamma(2\gamma_{1}+1)}
\nonumber \\
&& \times\left[\sum_{n=0}^{\infty}
\frac{\Gamma(n+\gamma_{2}-\gamma_{1}-1)
\Gamma(n+\gamma_{2}-\gamma_{1})}{n!\Gamma(n+2\gamma_{2}+1)}\right.
\nonumber \\
&& \quad
\left.+\sum_{n=0}^{\infty}\frac{\Gamma(n+\gamma_{2}-\gamma_{1}-1)
\Gamma(n+\gamma_{2}-\gamma_{1}+1)}{n!\Gamma(n+2\gamma_{2}+1)}\right].
\label{6}
\end{eqnarray}
Since it is known \cite[p.\ 36]{Magn66} that 
\begin{equation}
\sum_{n=0}^{\infty}\frac{\Gamma(n+a_{1})\Gamma(n+a_{2})}
{\Gamma(n+b)}\frac{z^{n}}{n!}
=\frac{\Gamma(a_{1})\Gamma(a_{2})}{\Gamma(b)}
\,{}_{2}F_{1}
\left(
\begin{array}{c}
a_{1}, a_{2} \\
b
\end{array}
;z
\right)
\qquad (|z|\leqslant1),
\label{7}
\end{equation}
where ${}_{2}F_{1}$ is the hypergeometric function, Eq.\ (\ref{6})
may be cast into the form
\begin{eqnarray}
\sigma_{+2} &=& \frac{2Z\alpha^{2}}{9}
\frac{\Gamma(\gamma_{2}+\gamma_{1}-1)\Gamma(\gamma_{2}+\gamma_{1}+2)}
{(\gamma_{2}-\gamma_{1})(\gamma_{2}-\gamma_{1}+1)
\Gamma(2\gamma_{1}+1)\Gamma(2\gamma_{2}+1)}
\nonumber \\
&& \times
\left[
{}_{2}F_{1}
\left(
\begin{array}{c}
\gamma_{2}-\gamma_{1}-1,\gamma_{2}-\gamma_{1} \\
2\gamma_{2}+1
\end{array}
;1
\right)
+(\gamma_{2}-\gamma_{1})
\,{}_{2}F_{1}
\left(
\begin{array}{c}
\gamma_{2}-\gamma_{1}-1,\gamma_{2}-\gamma_{1}+1 \\
2\gamma_{2}+1
\end{array}
;1
\right)
\right].
\nonumber \\
&&
\label{8}
\end{eqnarray}
The following identity \cite[p.\ 40]{Magn66}:
\begin{equation}
{}_{2}F_{1}
\left(
\begin{array}{c}
a_{1}, a_{2} \\
b
\end{array}
;1
\right)
=\frac{\Gamma(b)\Gamma(b-a_{1}-a_{2})}
{\Gamma(b-a_{1})\Gamma(b-a_{2})}
\qquad [\Real(b-a_{1}-a_{2})>0]
\label{9}
\end{equation}
is due to Gauss. Applying it to the two ${}_{2}F_{1}$ functions
appearing in Eq.\ (\ref{8}), we find
\begin{equation}
{}_{2}F_{1}
\left(
\begin{array}{c}
\gamma_{2}-\gamma_{1}-1,\gamma_{2}-\gamma_{1} \\
2\gamma_{2}+1
\end{array}
;1
\right)
=\frac{\Gamma(2\gamma_{1}+2)\Gamma(2\gamma_{2}+1)}
{\Gamma(\gamma_{2}+\gamma_{1}+1)\Gamma(\gamma_{2}+\gamma_{1}+2)}
\label{10}
\end{equation}
and
\begin{equation}
{}_{2}F_{1}
\left(
\begin{array}{c}
\gamma_{2}-\gamma_{1}-1,\gamma_{2}-\gamma_{1}+1 \\
2\gamma_{2}+1
\end{array}
;1
\right)
=\frac{\Gamma(2\gamma_{1}+1)\Gamma(2\gamma_{2}+1)}
{\Gamma(\gamma_{2}+\gamma_{1})\Gamma(\gamma_{2}+\gamma_{1}+2)}.
\label{11} 
\end{equation}
Inserting Eqs.\ (\ref{10}) and (\ref{11}) into Eq.\ (\ref{8}), after
some rearrangements involving, among others, the identity
\begin{equation}
\gamma_{2}^{2}=\gamma_{1}^{2}+3,
\label{12}
\end{equation}
we eventually arrive at the following expression for $\sigma_{+2}$:
\begin{equation}
\sigma_{+2}=\frac{2Z\alpha^{2}}{27}\frac{\gamma_{1}+2}{\gamma_{1}+1}.
\label{13}
\end{equation}
With no doubts, it looks much neater than the one in Eq.\ (\ref{3})!

Insertion of Eqs.\ (\ref{2}) and (\ref{13}) into Eq.\ (\ref{1}) leads
to the following closed-form representation of the magnetic dipole
shielding constant for the relativistic hydrogen-like atom in its
ground state:
\begin{equation}
\sigma=-\frac{2Z\alpha^{2}}{27}
\frac{4\gamma_{1}^{3}+6\gamma_{1}^{2}-7\gamma_{1}-12}
{\gamma_{1}(\gamma_{1}+1)(2\gamma_{1}-1)},
\label{14}
\end{equation}
which is identical with the expression found earlier by Moore
\cite{Moor99}, Pyper and Zhang \cite{Pype99} and Ivanov \emph{et
al.\/} \cite{Ivan09} (after it is taken into account that the latter
authors define $\sigma$ with the opposite sign).
\end{document}